\documentclass[12pt]{article}

\usepackage{scicite}

\usepackage{times}
\usepackage{graphicx}% Include figure files
\graphicspath{{figures/}}
\usepackage{dcolumn}% Align table columns on decimal point
\usepackage{bm}% bold math
\usepackage{xcolor}
\usepackage{epstopdf}
\usepackage{amsmath}
\usepackage{hyperref}
\usepackage[figurename=Fig.]{caption}
\usepackage[labelfont=bf]{caption}
\usepackage[labelsep=period]{caption}

\newenvironment{sciabstract}{%
\begin{quote} \bf}
{\end{quote}}

\title{Cluster formation in particle-laden flows is a continuous phase transition
}%\small{\textcolor{blue}{\\Complex networks built from particle-laden flows help identify local and collective behavior of particles in the flow.\\Teaser: (125 character 1 sentence summary for non-specialist readers) }}} 

\author{K Shri Vignesh,$^{1}$ Shruti Tandon,$^{1}$ Praveen Kasthuri,$^{1}$  R. I. Sujith$^{1\ast}$\\
\\
\normalsize{$^{1}$Department of Aerospace Engineering, Indian Institute of Technology Madras,}\\
\normalsize{Chennai 600036, India}\\
\\
\normalsize{$^\ast$To whom correspondence should be addressed; E-mail:  sujith@iitm.ac.in}
}

\date{}

\begin{document} 


\baselineskip24pt


\maketitle

%\paragraph{Complex networks approach identifies phase transition resulting from the local and global dynamics of inertial particles in the flow.
%Complex networks built from particle-laden flows help identify local and collective behavior of particles in the flow.

\begin{sciabstract}

%Studying particle-laden flows is essential to understand diverse physical processes such as rain formation in clouds, pathogen transmission and pollutant dispersal. Distinct clustering patterns are formed in flows with particles of different inertia (characterized by Stokes number $St$). We use, for the first time, complex networks to characterize both local and collective dynamics of particles. A giant cluster emerges in the derived network through a continuous phase transition as particles cluster into patterns in 2D Taylor-Green flow. Thus, we show that clustering in particle-laden flows can be viewed as a phase transition. The rate at which phase transition occurs is found to obey a universal behavior across various $St$ and a scaling-law is found between the transition time and $St$. Such findings identified by complex network approach can have important implications in understanding the rate of droplet size growth during rain formation in clouds.

Studying particle-laden flows is essential to understand diverse physical processes such as rain formation in clouds, pathogen transmission, and pollutant dispersal. Distinct clustering patterns are formed in such flows with particles of different inertia (characterized by Stokes number $\boldsymbol{St}$). For the first time, we use complex networks to study the spatiotemporal dynamics in such flows. We simulate particles in a 2D Taylor-Green flow and show that the network measures characterize both the local and global clustering properties. As particles cluster into specific patterns from a randomly distributed initial condition, we observe an emergence of a giant component in the derived network through a continuous phase transition. Further, the phase transition time is identified to be related to the Stokes number through a power law for $\boldsymbol{St < 0.25}$ and an exponential function for $\boldsymbol{0.25\leq St\leq 1}$. Our findings provide novel insights into the clustering phenomena in particle-laden flows.   %\textcolor{red}{The rate at which phase transition occurs is found to obey a universal behavior across various $\boldsymbol{St}$, and a scaling-law is found between the transition time and $\boldsymbol{St}$.}  %Such findings identified by complex network approach can have significant implications in understanding the rate of droplet size growth during rain formation in clouds. %Further, the phase transition time is identified to be related to the Stokes number through a power law for $\boldsymbol{St < 0.25}$ and an exponential function for $\boldsymbol{0.25\leq St\leq 1}$. Further, the phase transition time is identified to be related to the Stokes number through a power law when particles were gradually centrifuged out of and an exponential function when particles form caustics.

%Studying particle-laden flows is essential to understand disparate physical processes such as rain formation in clouds, pathogen transmission, and pollutant dispersal. Due to varying response times of different inertial particles (characterized by Stokes number), distinct clustering patterns are formed in particle-laden flows. Here, we propose the use of particle proximity networks, to analyse the clustering characteristics for different Stokes number in a canonical 2D Taylor-Green flow. Both local and collective behavior of particles can be characterized during clustering using this network approach. We show that, a giant cluster forms through continuous phase transition in the network as pattern formation occurs in the particle-laden Taylor-Green flow. Moreover, we identify a scaling between the time-interval of phase transition and particle Stokes number. Our findings can have important implications in understanding timescales associated with collective behavior of droplets in cloud formation or spread of pathogen on sneezing.

\end{sciabstract}

\section*{INTRODUCTION}

Particle-laden flows are ubiquitous across diverse physical phenomena \cite{shaw2003particle,sahu2018interaction,mittal2020flow,haszpra2011volcanic,smith2002significance}. The underlying flow field and the inertia of particles play a crucial role in determining the individual as well as the collective dynamics of the dispersed particles in particle-laden flows. For example, understanding the rate of collision and coalescence between water droplets is essential to understand the process of rain formation in clouds \cite{bodenschatz2010,grabowski2013growth,pumir2016collisional}. In wind-pollinated plants, the success rate of reproduction depends on the diffusion and transport characteristics of pollen grains in wind \cite{sabban2011measurements}. Furthermore, clustering amongst inertial particles in flows significantly affects the rate of collision between them \cite{pumir2016collisional, grabowski2013growth}. In droplet combustion, the formation of droplet clusters has a significant effect on the combustion process \cite{chiu1977group}. Of more immediate relevance is the need to develop an understanding of how virus-laden respiratory droplets transmit from a host to uninfected recipients, for example, during the transmission of COVID-19 \cite{mittal2020flow}.

In all the above scenarios, the dispersed phase (particle) is denser than the continuous phase (the fluid medium). The inertia of a particle is characterized by its Stokes number ($St$), which determines the response time of the particle ($t_p$) to changes in the flow. For a particle of density $\rho$, dynamic flow viscosity $\mu$, and particle radius $a$, $t_p$ is given by, $ t_{p} = 2\rho_{p}a^2/9\mu$ \cite{crowe1998multiphase}. Stokes number ($St$) is a non-dimensional ratio of the particle response time $t_{p}$ to the time scale of fluid flow $t_{f}$. Inertial particles are more likely to cluster in regions of high shear than in regions of high vorticity, owing to particles being centrifuged out of the vortex region \cite{squires1991preferential, eaton1994preferential}. The tendency of inertial particles to cluster in the shear region is found to be higher when $St \sim O(1)$ \cite{wang1993settling, eaton1994preferential, squires1991preferential, chun2005clustering}. 

%\textcolor{blue}{
%Several attempts have been made to charaterize clustering in particle-laden flows. Aliseda \textit{et al.}\cite{aliseda2002effect} studied the clustering characteristics of water droplets in turbulence using box counting method. Using RDF method, Saw \textit{et al.}\cite{saw2008inertial} observed the clustering amongst inertial particles to be geometrically self-similar. Both box-counting method and RDF method provide only a global measure of the clustering amongst particles\cite{monchaux2012analyzing}, but are unable to explain individual particle behavior. For a better understanding of the motion of local dynamics, Lagrangian information on the local clustering properties is very important\cite{monchaux2012analyzing}. Models on the rate of collisions between particles are developed with the help of local clustering characteristics of particles \cite{pumir2016collisional}. Monchaux \textit{et al.}\cite{monchaux2010preferential} used Voronoi diagrams to study the local clustering characteristics. In this method, the flow domain is divided into cells such that each cell houses a single particle. The local clustering characteristics of a particle is inversely proportional to the area of the cell. The results are however highly sensitive to the number of particles in the flow domain\cite{tagawa2012three}. As the clustered regions are identified by a threshold on the size of the cell, Voronoi analysis can fail to capture the local clusters amongst particles. }\textcolor{red}{(@vignesh will shorten )}

Currently, several approaches are used to characterize the clustering of inertial particles in fluid flows. The box-counting method\cite{aliseda2002effect} and radial distribution function method\cite{saw2008inertial,sundaram1997collision,lian2013preferential} provide a global measure of the clustering; however, they are unable to provide a Lagrangian (local) information of clustering amongst particles. Local clustering characteristics are essential to better understand the dynamics of individual particles and their collective behavior, which is significant for developing models on the rate of collisions between particles \cite{pumir2016collisional}. So far, only Voronoi diagrams are able to characterize and quantify the local clustering characteristics\cite{monchaux2010preferential}. However, the results obtained using this method are sensitive to the total number of particles in the flow domain \cite{tagawa2012three}. Hence, Voronoi diagrams can sometimes fail to capture the local clustering characteristics among particles. 

Fluid flows laden with inertial particles are essentially complex systems. Complex systems consist of interacting subsystems whose collective behavior cannot be deduced by studying each of the subsystems in isolation. Collective behavior by the constituents of a complex system often leads to interesting phenomena such as synchronization, phase transition, and the emergence of order in the system\cite{gao2017complex,iacobello2020review}. The behavior of a typical particle-laden flow is controlled by the complex nonlinear interactions between its constituent subsystems such as flow structures (vortex regions, shear regions) and inertial particles \cite{bormashenko2020clustering, brandt2021particle}. The linear and nonlinear interactions between the flow structures and particles lead to the formation of spatiotemporal patterns and particle clusters. The organization of particles into clusters is essentially a manifestation of the emergence of order in this complex system.

The natural language for describing complex systems is that of complex networks. A network is formed by nodes representing the constituents of a system, joined by edges representing their interactions \cite{barabasi2016network}. Complex networks have been extensively used in studying spatiotemporal dynamics of climate systems \cite{scarsoglio2013climate},  geophysical flows\cite{ser2015flow}, turbulence characterization \cite{iacobello2018spatial, taira2016network} and in thermo-fluid systems \cite{unni2018emergence,iacobello2020review}. Several methods of network construction involving correlation and event synchronization between dynamic variables or using Lagrangian interactions, proximity, and visibility algorithms have been proposed \cite{iacobello2020review}. We propose the use of complex networks \cite{barabasi2016network} to characterize the spatial inhomogeneities arising as a result of inertial particle dynamics in fluid flows.

Here, we build proximity-based networks for particle-laden flows where the particles are nodes. The connections between these particles are established based on the local distribution of particles in the neighbourhood of a chosen particle, derived using the radial distribution function \cite{sundaram1997collision,lian2013preferential}. We consider a particle-laden 2D Taylor-Green flow where the underlying flow is characterized by symmetric vortical structures separated by shear regions \cite{taylor1937mechanism}. The particle dynamics is modeled using the Maxey-Riley equation \cite{maxey1983equation,maxey1987motion} which is a simple force balance between the viscous drag force and the particle inertia. 

To the best of our knowledge, ours is the first application of complex networks to understand the dynamics of inertial particles. Further, the proposed method facilitates the characterization of clustering amongst individual particles while simultaneously providing a global measure of the overall spatial distribution of particles. Local network measures such as the degree of nodes help us infer the clustering characteristics of individual particles. In contrast, the global clustering of particles is characterized using global measures such as average clustering coefficient and size of the spanning cluster. As clusters of particles are formed in the flow, a giant cluster emerges in the corresponding network via a continuous phase transition. We thus propose that clustering in particle-laden flows is a phase transition. This perspective is further used to derive novel insights about the rate of clustering in particle-laden flows. 

Complex network analysis and network-based-models have been successfully used to characterize phase transitions in disparate real-world systems \cite{dorogovtsev2008critical,achlioptas2009explosive,lee2018recent}. For example, the occurrence of systemic risk in financial networks is characterized by a phase transition in a network of agents bound by game-theoretic rules \cite{elliott2014financial,gai2010contagion}. Epidemic spread is seen as a phase transition similar to Bose-Einstein condensation on a network \cite{tang2009influence}. The onset of oscillatory instabilities in thermo-fluid systems can be viewed as a phase transition \cite{tandon2021condensation}. Phase transitions in complex networks, often referred to as percolation transitions, have also been investigated using various mathematical models \cite{erdos1960evolution,achlioptas2009explosive}. The transition can either be continuous (second-order transition), or it may be abrupt (discontinuous or first-order transition). Both continuous and abrupt (often called explosive) percolation transitions have been reported for complex networks using various mathematical models \cite{dorogovtsev2008critical,d2019explosive}.% However, the debate on whether explosive transitions in networks are indeed discontinuous is still on \cite{riordan2011explosive}.

In our system, during phase transition, we observe randomly distributed particles organizing into patterns and clusters due to their inertia and the underlying flow field. By analyzing the phase transition in the derived networks, we characterize the duration of clustering in particle-laden Taylor-Green flows for various $St$. The insight so developed can have important implications on explaining the dispersal time of pathogens in the atmosphere or the rate of increase of droplet size in clouds with variation in the $St$. As a paradigm for analyzing inertial particle dynamics, the proposed approach proves to be a potential tool to resolve various questions regarding particle-laden flows.

\section*{RESULTS}
\subsection*{ Constructing particle proximity networks }

Numerical simulations for particle-laden 2D Taylor-Green flow are performed for various $St$ in the range $0.01\leq St \leq 3$. The particle dynamics is modeled using the Maxey-Riley equation (see materials and methods for the model). 

To obtain the clustering characteristics from the spatial distribution of particles, we consider individual particles as nodes of the network. The adjacency matrix of the network has $N \times N $ elements, where $N$ is the number of particles in the flow domain.  Particle positions at a time instant from the simulation form the basis for our network construction. The link between the particles is established based on the proximity determined from the radial distribution function ($g(r)$), which provides a statistical measure on the spatial distribution of particles in the neighbourhood of the particle in consideration \cite{sundaram1997collision,lian2013preferential} and is calculated using the following equation:

\begin{equation}\label{eq5}
g(r) = \frac{N_r(r)}{\delta A} \bigg/ {\frac{N}{A}}. 
\end{equation}
Here, $N_r$ is the number of particles found within a radial distance $r$ from the reference particle, $\delta A$ represents the differential area element of thickness $dr$ constructed at a distance $r$ from the reference particle. $N$ is the total number of particles in the flow domain and $A$ is the area of the flow domain. When particles are uniformly distributed around the reference particle, $g(r) = 1$. However, if the particles are clustered in the vicinity of the reference particle, $g(r) > 1$ and if the particles are sparsely distributed, $g(r) < 1$. We calculate $g(r)$ for all the particles in the flow domain. From the calculated $g(r)$, we find the critical radius $r_c$, the first intersection point of $g(r)$ and $g(r) = 1$, with $dg/dr \Big|_{r=r_c}< 0$. Links are established between two particles if any one of the particle is within the radius $r_c$ of the other particle.

\begin{figure}[t!]
\includegraphics[width=1\textwidth]{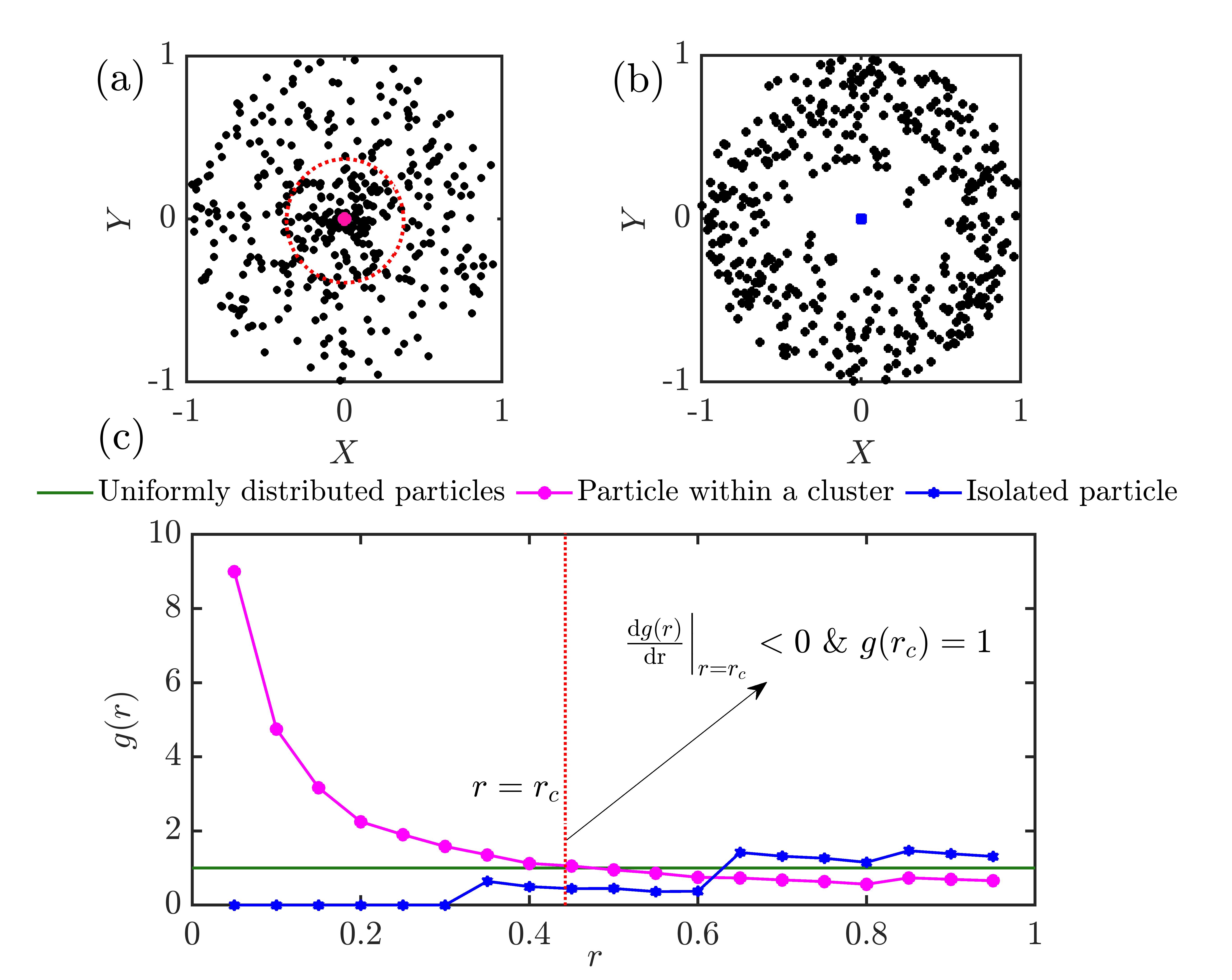}
\caption{\textbf{Radial distribution function ($\boldsymbol{g(r)}$) captures the spatial distribution of particles relative to the reference particle}. Spatial distribution of particles illustrating (a) particle within a cluster and (b) an isolated particle. The dotted circle marks the critical radius ($r_{c}$) for the clustered particle. (c) The corresponding $g(r)$ variation for the clustered and isolated particle.}
%Cluster radius ($r_c$) for the clustered particle is estimated as $g(r_c) = 1$ and $dg/{dr} \big|_{r=r_c}< 0$}. Particles within the radius $r_c$ are connected to the reference particle in the network.
\label{gr_plot}
\end{figure}

Let us consider two scenarios to illustrate the use of $g(r)$ in our network construction, a particle that is well entrenched in a cluster (Fig.~\ref{gr_plot}a), and an isolated particle where particles are sparely distributed in its vicinity (Fig.~\ref{gr_plot}b). The variation of $g(r)$ for both the particles is shown in Fig.~\ref{gr_plot}c. Here, for the clustered particle, $g(r)$ is greater than one in its vicinity and decreases in the local neighborhood along the radial direction based on the spatial distribution of particles. Particles within the radius $r_c$ (see Fig.~\ref{gr_plot}a ) are clustered relative to the reference particle and hence are connected to the reference particle in the network. For the isolated particle (see Fig.~\ref{gr_plot}b), $g(r)<1 $ in its vicinity and increases beyond one after a specific radial distance. From the variation of $g(r)$, we can infer that though there are particles in the neighborhood of the reference particle, they are not in the immediate vicinity. Hence, no connection is established with this isolated particle. Following this approach, the existence of a link between any two particles is encoded into a binary adjacency matrix ($A_{ij}$) as 1 if they are connected or 0 otherwise. The links established between the particles are unweighted and undirected. Since we do not consider any self-connections, $A_{ii}$ is set to 0.

Using network measures such as the degree of a node ($k$) and the local clustering coefficient ($C$), we characterize the clustering between particles. The degree of a node ($k$) quantifies the number of connections a node has with other nodes in the network. Higher $k$ indicates the presence of many particles in the vicinity of a node. To remove the dependence on the number of particles in the flow domain, we normalize $k$ with the total number of nodes in our system. The average normalized degree ($\langle k \rangle$), obtained by averaging the degree across all the nodes in the network (Eq.~\ref{eq9}), provides a global estimate on the clustering amongst particles. For studying the changes in the network structure, we use degree distribution ($P(k)$ vs. $k$), where $P(k)$ is the probability of a randomly selected node having degree $k$.

The local clustering coefficient ($C$) captures the inter-connectivity between the neighbours of the reference node. $C$ is a local measure useful in capturing the cluster density between the neighbours of the reference node. A particle entrenched in a densely concentrated region of a cluster will have a higher value of $C$ than a particle in a moderately concentrated region. The value of $C$ lies between 0 and 1, where 0 corresponds to an absence of links between neighbors and 1 corresponds to all the neighbors being connected to each other. For obtaining a global estimate on how densely the particles are clustered, we average the clustering coefficient ($\langle C \rangle$) across all the nodes in the network.

\subsection*{Network analysis of clustering in particle-laden 2D Taylor-Green flow}

\begin{figure}[h!]
\centering
\includegraphics[width=1\textwidth]{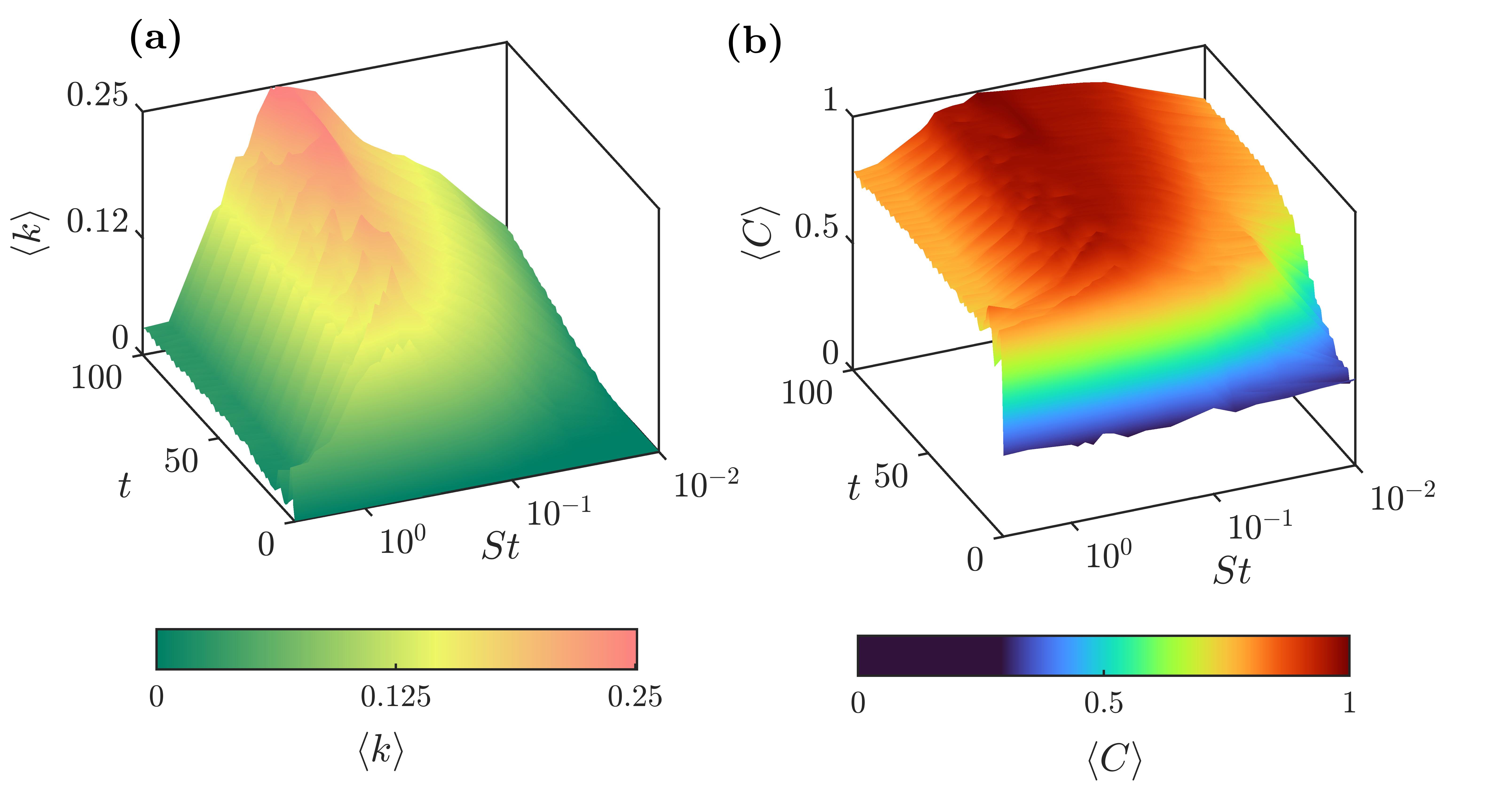}
\caption{ \textbf{Rise and saturation of $\boldsymbol{\langle k\rangle}$ and $\boldsymbol{\langle C \rangle}$ quantify the formation of clusters.} Temporal variation of $\langle k \rangle$ in (a) and $\langle C \rangle$ in (b) for particles with $St$ in the range $0.01$ to $3$. The dependence of the rate of clustering on $St$ is captured by $\langle k \rangle$ and $\langle C \rangle$. }
\label{Avg_plot}
\end{figure}

From the spatial distribution of particles at each time instant, a proximity-based complex network is constructed and characterized using $\langle k \rangle$ and $\langle C \rangle$. As particles cluster in time, $\langle k \rangle$ and $\langle C \rangle$ of the networks increases. When particles asymptotically settle into the stagnation points, the values of $\langle k \rangle$ and $\langle C \rangle$ saturate at $0.25$ and $1$, respectively. Weakly inertial particles ($St<0.1$) cluster slowly which is evident from the gradual increase of $\langle k \rangle$ and $\langle C \rangle$ (Fig. \ref{Avg_plot} and movie S1). 

\begin{figure}[h!]
\includegraphics[width=1\textwidth]{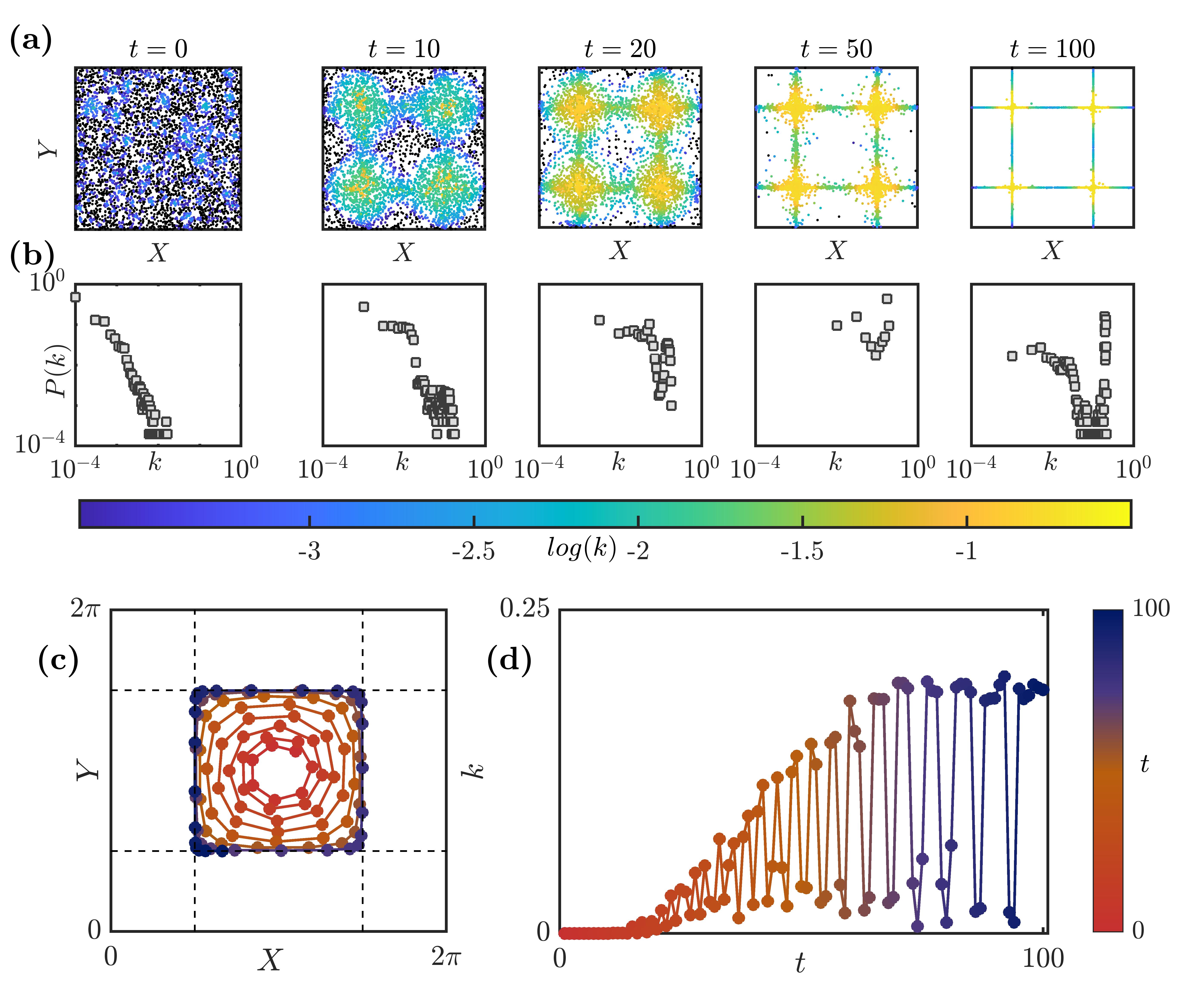}\caption{\textbf{ Particles having $\boldsymbol{St=0.05}$ cluster slowly due to their weak inertia.} Spatial distribution of the particles in the flow domain color coded based on its degree ($k$) in log scale. Particles without any connections are identified by black color. (a) Increase in $k$ of particles and (b) changes in the initial power-law degree distribution ($P( k)$ versus $k$) with time as particles form clusters that resemble a cross in the shear regions for the flow field. (c) The trajectory of a particle originating inside a vortex region does not cross the separatrix in the shear region (identified by the broken lines). (d) The slow centrifugal motion of the particle is reflected in the variation of $k$, which characterizes the instantaneous local clustering. Here, the oscillations indicate the movement of the particle towards and away from the stagnation points. }
\label{St005}
\end{figure}

\begin{figure}[h!]
\centering
\includegraphics[width=1\textwidth]{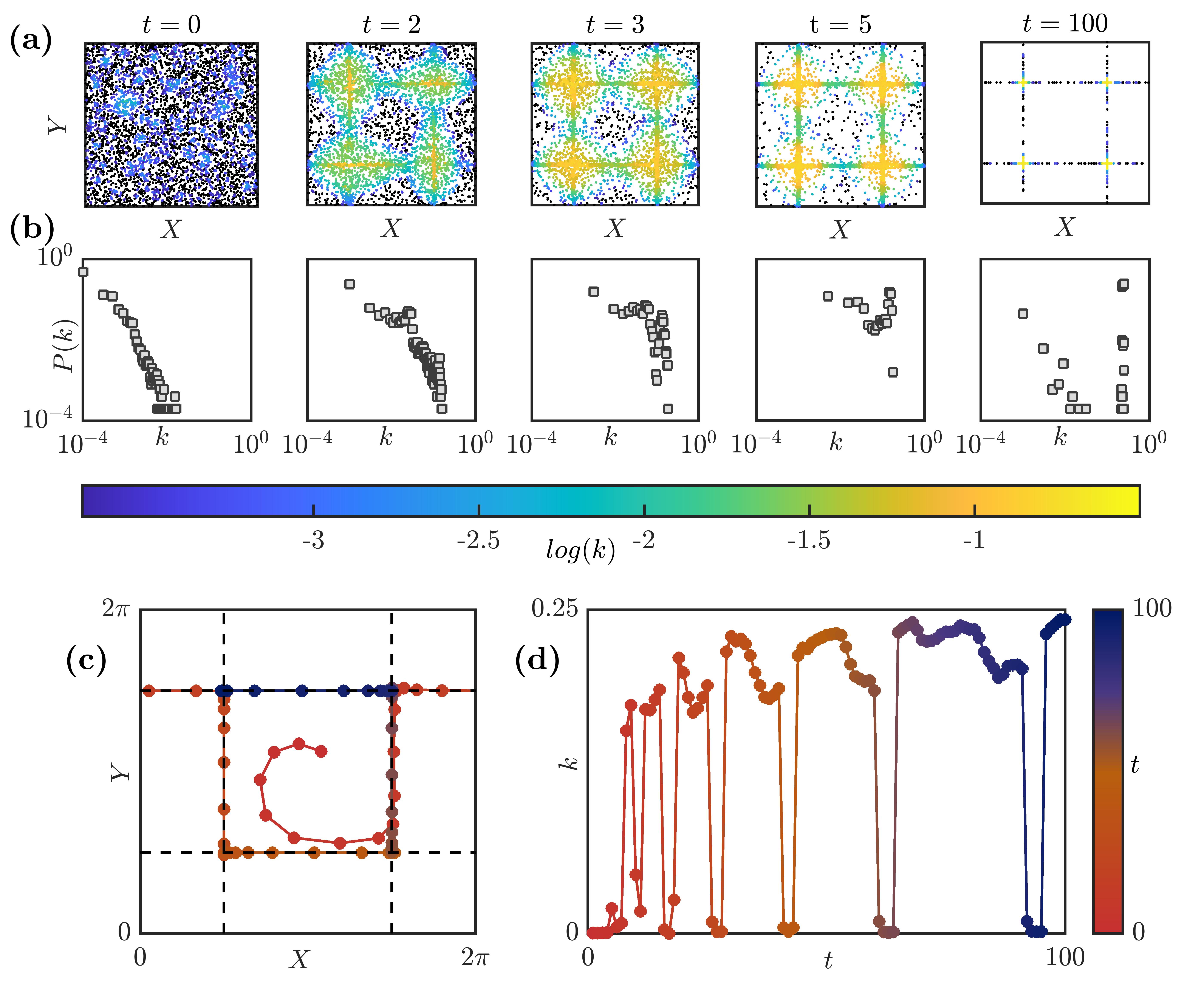}\caption{\textbf{Rapid centrifuging of $\boldsymbol{St=0.5}$ particles swiftly form cluster patterns in shear regions.} Spatial distribution of the particles in the flow domain color coded based on its degree ($k$) in log scale. Particles without any connections are identified by black color. (a) Increase in $k$ of particles and (b) distortion of the  initial power-law degree distribution ($P( k)$ versus $k$) with time as particles cluster. (c) Due to sufficient inertia of $St = 0.5$ particle, the trajectory of a particle originating inside a vortex crosses the separatrix (identified by the broken lines). (d) The degree ($k$) of the particle captures the instantaneous local clustering.  The swift increase in $k$ indicates the rapidness of the particle being centrifuged out, and the oscillations reflect the motion between the stagnation points.}
\label{St05}
\end{figure}

\begin{figure}[h!]
\includegraphics[width=1\textwidth]{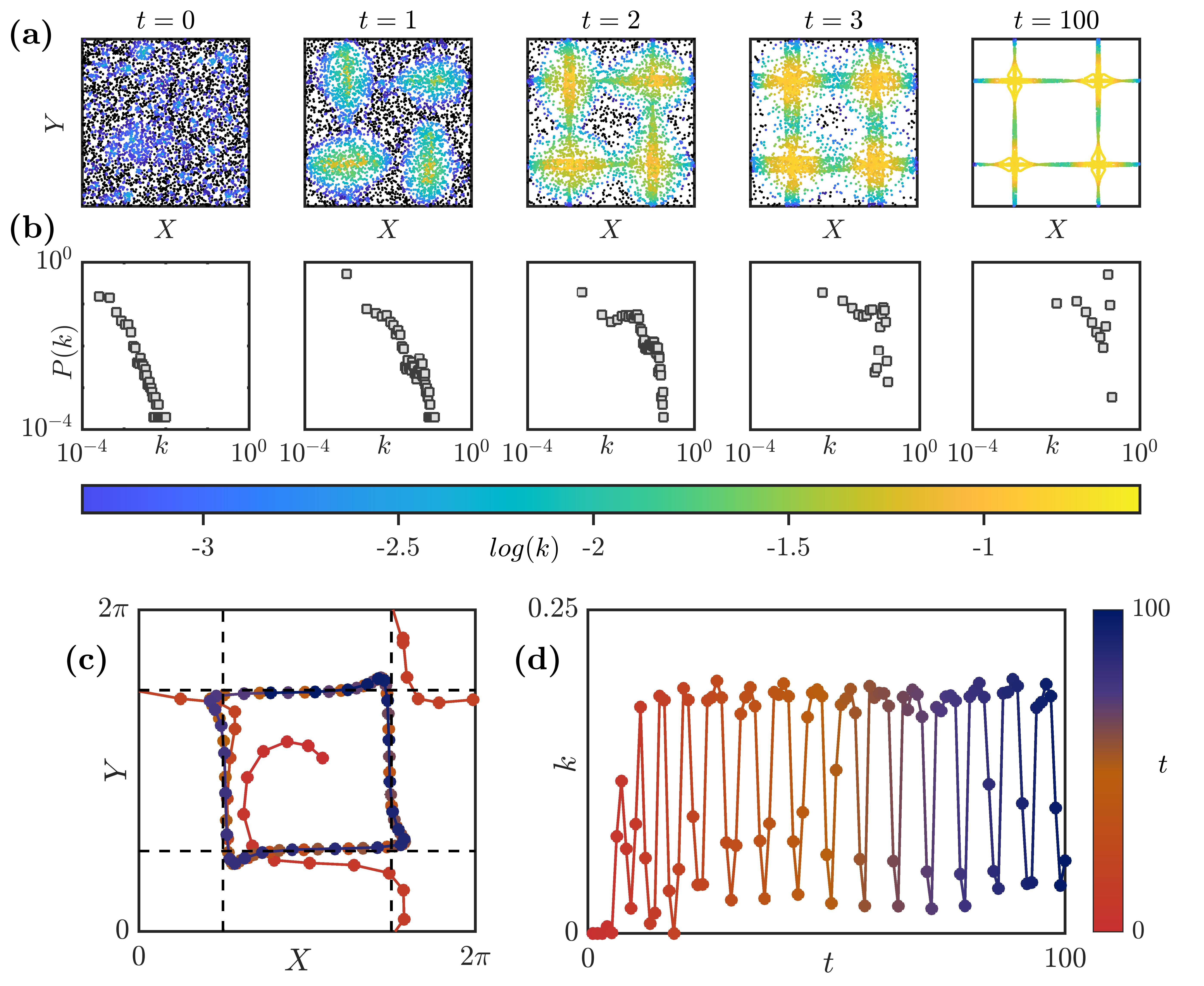}
\caption{\textbf{Rapid centrifuging and formation of patterns that resemble a cross as particles of $\boldsymbol{St = 1}$ cluster in shear regions.}  Spatial distribution of the particles in the flow domain color coded based on its degree ($k$) in log scale. Particles without any connections are identified by black color. (a) Increase in $k$ of particles and (b) distortion of initial power-law degree distribution ($P( k)$ versus $k$) with time as particles form cross-patterns and clusters in the flow. (c) The trajectory of a particle originating inside a vortex region crosses the separatrix (identified by the broken lines) and settles into a periodic orbit. The oscillation in $k$ of such a particle reflect the motion of the particle trapped in the periodic orbit.}
\label{St1}
\end{figure}

As we increase $St$, the rate at which particles are centrifuged out increases \cite{ravichandran2015caustics}; hence, for $St \sim O(1)$, both $\langle k \rangle$ and $\langle C \rangle$ values increase rapidly. For $St \leq 0.7$, all particles in the flow domain asymptotically settle into the stagnation points (movies S1 and S2). 
Hence, as $t \rightarrow 	\infty$, both $\langle k \rangle$ and $\langle C \rangle$ saturate at $0.25$ and $1$ respectively. For $St > 0.7$, a fraction of particles exhibit periodic motion without settling into the stagnation points \cite{nath2021transport} (movie S3). As we increase the $St$, the fraction of particles exhibiting periodic motion increases. Hence, we observe a decrease in the saturated values of $\langle k \rangle $ and $\langle C \rangle$ for $St > 0.7$. Furthermore, as a result of the periodic motion of particles, we observe undulations in the saturated $\langle k \rangle$ and $\langle C \rangle$ (Fig. \ref{Avg_plot}). For $St > 1$,  all the particles exhibit periodic motion without clustering in the shear regions, which is reflected in the lower values of $\langle k \rangle$ and $\langle C \rangle$. Thus, the emergence of clusters for $0.01\leq St \leq 1$ saturates the average network measures at a high value reflecting the global clustering characteristics. These observations are illustrated and analyzed here in detail for the specific cases of $St=0.05$, $St=0.5$ and $St=1$ as shown in Fig. \ref{St005}- \ref{St1}. 

%%%%%%%%%%%%%%%%%%%%%%%%%%%%%%%%%%%%%%%%%%%%%%%%%
We start with a random initial distribution of particles in the flow domain (Fig. \ref{St005}-\ref{St1}(a) at $t=0$). In the corresponding network at $t=0$, most particles have a few connections ($k\approx 0$) while a few nodes have more than just a few connections ($k\approx0.01$). Corresponding, we observe a power law behaviour in the degree distribution for randomly distributed particles (Fig. \ref{St005}-\ref{St1}(b) at $t=0$). The inertia of particles comes into play for $t>0$, and the simulations are performed up to $t = 100$. 

As we progress in time, particles move out of their streamlines and form prominent cluster patterns. The time at which cluster patterns become apparent differ with $St$; for $St=0.05, 0.5$ and $1$, patterns that resemble a cross emerge in the shear regions at $t = 20, 3$ and $2$, respectively (movies S1, S2 and S3). The scatter plots in Fig. \ref{St005}-\ref{St1}(a) show the increase in the degree of particles that start clustering in the shear region. Correspondingly, we observe a distortion of the original power-law degree distribution in Fig. \ref{St005}-\ref{St1}(b). As particles form dense clusters, the number of particles with higher $k$ increases significantly. 
Thus, the corresponding network possesses a large fraction of nodes with higher degrees. In the degree distribution at $t = 100$, the particles with high $k$ correspond to the particles near the stagnation points while particles with low $k$ correspond to the few particles commuting between these stagnation points in the shear regions. Finally, at very large $t$, for $St = 0.05$ and $0.5$, all particles settle into the four stagnation points, thus exhibiting a steady state clustering (movies S1 and S2).% However, for $St > 0.7$, all the particles do not settle into any stagnation points. 

The trajectory of a particle initiated in the vortical region is governed by its $St$ and the underlying flow field. Particles with $St\leq0.25$, do not cross the separatrices in the shear regions due to their weak inertia\cite{healy2005full,nath2021transport}. Instead, the particle gets slowly centrifuged outwards and asymptotically settles into a stagnation point in the edge of the vortex (see Fig.\ref{St005}c ). As this particle starts nearing the stagnation point, the number of particles in its neighbourhood start to increase and hence its $k$ value increases. The oscillations in $k$ arise as the particle approaches and moves farther from stagnation points due its the centrifugal motion. The peaks and troughs in $k$ occur when a particle is nearest and farthest from the stagnation points respectively. Moreover, the frequency of the oscillations in $k$ are directly related to the frequency of the spiral motion of the particle in the vortical region. 

On the other hand, a particle with $St>0.25$ initialized inside a vortical region has sufficient inertia to cross the separatrices. Thus, particles originating in different regions can interact with each other \cite{healy2005full,nath2021transport} (evident in Fig. \ref{St05}(c)). Since a particle in the vortical region is quickly centrifuged out and enters the shear region, $k$ of the particle increases sharply. The frequency of oscillation in $k$ is determined by the time the particle takes to move from one stagnation point to the other in the shear region.

Finally, for $St=1$, the particle response time matches with the flow time scale. Particles are ejected very rapidly from the vortex regions and form dense clusters in the shear regions (Fig. \ref{St1}(a) and movie S3). As a result, $k$ of the particle increases sharply. These particles do not asymptotically settle into the stagnation point; instead, they move in specific periodic orbits in the shear region (Fig. \ref{St1}(c)). The frequency of oscillation of $k$ is clearly a function of the frequency of particle motion in its periodic orbit. The patterns formed by a collection of such particles remain consistent with time (Fig. \ref{St1}(a) at $t = 100$). %It is noteworthy, that the frequencies of variation of $k$ of particles with $St=0.05, 0.5$ and $1$ are different owing to the influence of $St$ in determining the trajectory of the particle in each case.

\subsection*{Clustering as a continuous phase transition}\label{phase_transition}

%The organization of particles into patterns that resemble a cross leads to formation of \textcolor{red}{large number of connections} in the derived proximity networks.

The connections between the nodes in the network increase drastically as particles organize into cluster patterns that resemble a cross. The size of the largest component $(C_M)$ is defined as the total number of nodes in the largest connected set of nodes in the network \cite{Larremore2021}. Here, we draw inferences about the process of clustering in particle-laden 2D Taylor-Green flow by studying the variation of $C_M$ of the network, which indicates the extent of clustering amongst particles at any given time. The normalized size of the largest component ($C_M/N$) changes drastically from $(C_M/N) \sim 0$ to $(C_M/N) \sim 1$ over a finite time interval ($\Delta t$) characterizing a phase transition. In this study, when $(C_M/N) \sim 1$ , the largest component is referred to as the giant component. The emergence of a giant component in the network often occurs via a phase transition \cite{albert2002statistical}. Here, we use $C_M$ as the order parameter to characterize this phase transition.

\begin{figure}[h!]
\centering
\includegraphics[width=0.8\textwidth]{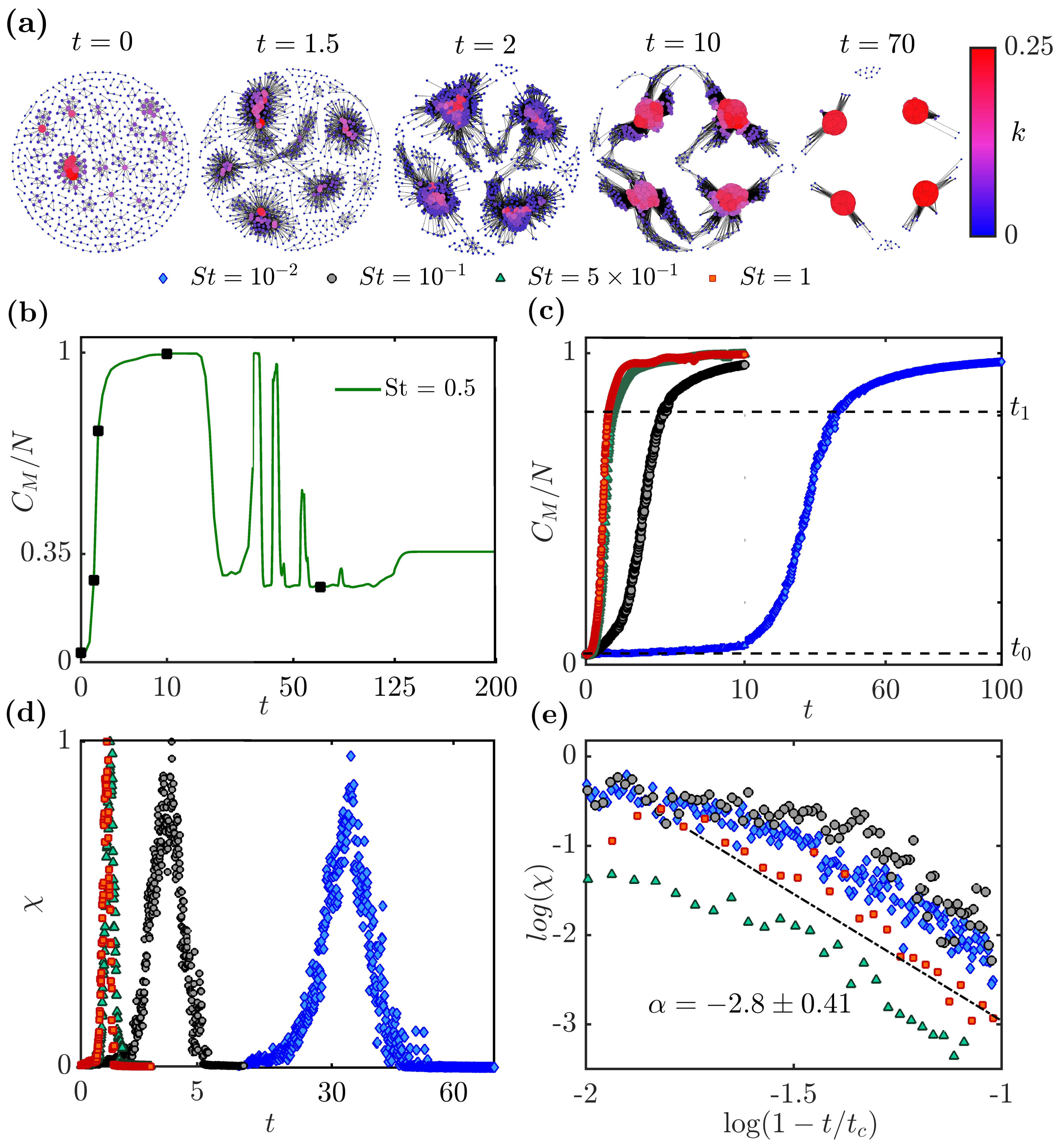}
\caption{\textbf{A continuous phase transition occurs in the derived proximity-based networks as particles cluster in the flow field.} Formation of cluster patterns that resembles a cross, increases the size of the largest component leading to the formation of a giant component. The formation of a giant component can be seen from the changes in the network topology from $t = 0$ to $t = 10$ (a) for $St = 0.5$ particles visualized using the Fruchterman-Reingold algorithm \cite{fruchterman1991graph} in the Gephi visualization software \cite{bastian2009gephi}. Variation of the normalized size of the largest component of network ($C_M/N$) with time, where the markers indicate $t = 0, 1.5, 2, 10, 70$ respectively. Comparison of the temporal variation of (c) the size of the largest component and (d) divergence of susceptibility ($\chi$) around critical time, for different $St$. (e) The variation of $\chi$ expressed as a function of critical time ($t_c$) collapses to a single curve on log-log scale highlighting universality in the phase transition with a critical exponent $\alpha = - 2.8\pm 0.41$ with $90\%$ confidence.}
\label{PT}
\end{figure}

For particles with $St=0.5$, the phase transition in the derived networks occurs around $t=2$ (see Fig.~\ref{PT}a-b) as $C_M$ increases rapidly in a short epoch owing to the rapid clustering of particles in the shear regions (see Fig. ~\ref{St05}a). The results in Fig.~\ref{PT} are obtained for an ensemble of $200$ realizations and for a time resolution of 0.02. The size of the giant component saturates and reaches a maximum for a significant epoch between $t=5$ to $10$ (see Fig.~\ref{PT}b). During this epoch, a significant proportion of particles commute in the shear regions between the stagnation points (see Fig.~\ref{St05}a). As particles asymptotically settle into the stagnation points, the number of particles commuting between stagnation points in the shear region reduces leading to the fragmentation of the network and a corresponding decrease in $C_M$. Note that, the number of particles that settle asymptotically into each stagnation point is not necessarily equal after complete clustering occurs in the flow field, and thus the size of the largest cluster not necessarily tends to $N/4$ as $t \to \infty$ (see Fig.~\ref{PT}b). The oscillations in $(C_M/N)$ occur due to the decrease in connections in the network as the number of particles commuting between stagnation points reduces. %Note that, the number of particles asymptotically settled into each stagnation point is approximately equal, and therefore $C_M$ approximately tends to $N/4$ as $t \to \infty$. The oscillations in $(C_M/N)$ occur due to the loss of connections in the network as the number of particles commuting between stagnation points reduces.

%For $St=0.5$ particles, a phase transition occurs around $t=2$ (Fig. \ref{PT}a) as the size of the largest component increases rapidly in a short epoch owing to the rapid motion of particles into the shear regions (Fig. \ref{St05}a). The results in Fig. \ref{PT} are obtained for an ensemble of $200$ realizations and for a time resolution of 0.02. \textcolor{red}{Note that after this phase transition, the pattern of particles alters dynamically (see variation of pattern in Fig. \ref{St05}(a) between $t=2$ to $t=5$), and almost all particles are in motion between the stagnation points.} However, the network properties saturate (Fig. \ref{Avg_plot}) and so does the size of the spanning cluster around $t=2$ for a significant epoch between $t=5$ to $10$ (Fig. \ref{PT}). After $t=20$, some particles come to a halt at the stagnation point and velocity of some other particles begin to decrease (Fig. \ref{St05}). These particles settle asymptotically into the $4$ stagnation points in the selected flow field, and as a result the maximum cluster size decreases. Note that, the number of particles trapped in each stagnation point is not necessarily equal after complete clustering occurs in the flow field, and thus the size of the spanning cluster tends approximately, and not necessarily, to $N/4$ as $t \to \infty$. The oscillations in $\mathcal{O}(C_M)$ occur primarily due to variation in the number of particles commuting between stagnation points at any given time. 

\begin{figure}[h!]
\centering
\includegraphics[width=0.7\textwidth]{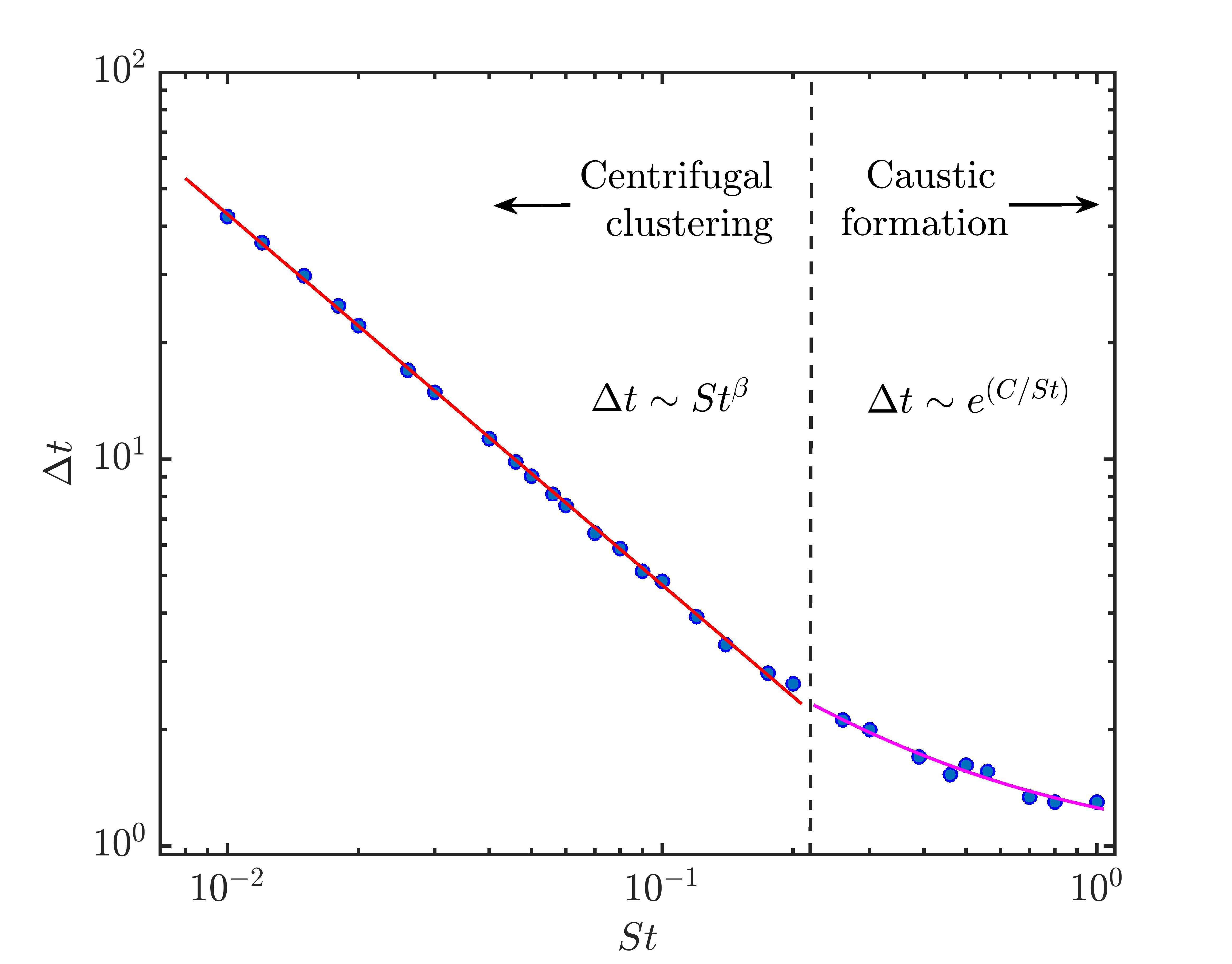}
\caption{\textbf{Variation of the time interval of phase transition $\boldsymbol{\Delta t}$ for different regimes of clustering.} $\Delta t$ varies as a power-law with scaling exponent  $\beta = -0.95\pm0.01$ with $90\%$ confidence for $St<0.25$ and varies as exponential for $0.25\leq St \leq 1$ with a constant $C = 0.18\pm0.02$ with $90\%$ confidence.}
\label{fig_deltaTwithSt}
\end{figure}

The rate of phase transition in proximity networks is a strong function of $St$ (Fig.~\ref{PT}(c)).
Significant variation in the $(C_M/N)$ observed in Fig.~\ref{PT}(c) occurs precisely when patterns that resemble a cross become apparent in the flow (Fig.~(\ref{St005}-\ref{St1})a). The time interval of phase transition ($\Delta t$) is determined by setting thresholds for the order parameter ($C_M$). We define $\Delta t = t_1 -t_0$, where $t_0$ is the maximum time at which $C_M \leq N^{\gamma} $, i.e., when $C_M$ is a fractional power of the network size. Further, $t_1$ is the minimum time at which $C_M \geq AN$, i.e., when $C_M$ is a significant fraction of the network size. For the results presented in our analysis, we choose $A = 4/5$ and $\gamma = 3/5$. We observe that $\Delta t$ decreases as we increase $St$ from $0.01$ to $1$ (Fig.~\ref{fig_deltaTwithSt}). Clearly, $\Delta t$ depends on the time taken for the particles to form clusters of discernible patterns in the flow.

A typical feature of phase transitions is the divergence of the variance of fluctuations of the order parameter, called susceptibility ($\chi$), around the critical time when the transition occurs. Susceptibility ($\chi$) is defined as $<C_M^2> - <C_M>^2$, where $< >$ is the ensemble average operator. From Fig. \ref{PT}(d), we observe that susceptibility diverges at faster rate and at lower critical times ($t_c$) for higher values of $St$. Here, $t_c$ is identified as the time at which $\chi$ attains its peak value in Fig. \ref{PT}(d). For lower values of $St$, this transition occurs over a longer time interval $\Delta t$ and at higher values of $t_c$ when compared to that during higher values of $St$. Shorter epochs of phase transition for particles with higher $St$ indicate the tendency of such particles to cluster rapidly. In Fig. \ref{PT}(e), we plot the variation of $\chi$ as a function of $(1 - t/t_c)$ on a log-log scale. Firstly, we note in Fig. \ref{PT}(e) that the critical behavior during phase transition for various $St$ exhibits a power-law scaling around the critical time as $\chi = (1 - t/t_c)^{\alpha}$, with critical exponent $\alpha = -2.8\pm 0.41$ with $90\%$ confidence. % Note that, while critical transition time ($t_c$) and rate of transition ($\Delta t$) differ significantly for various $St$, the critical exponent ($\alpha$)  variation of susceptibility in the vicinity of $t_c$ for different $St$ collapses to the same curve, indicating an underlying universality during collective organization of inertial particles of different $St$ into clusters.} 

%@vignesh what is the value of gamma, A beta for these cases? how many ensembles were used? what other necessary details? everything you needed for simulation should be written... modify the numbers once you have done more calculations...also mention the time resolution in captions of all figs in this section and also mention in text somewhere all necessary details of the simulation like ensemble and temporal resolution and till what time is the simulation done
%To further investigate this universality in scaling of the order parameter across several $St$ close to the transition point, we study the effect of varying the thresholds on the order parameter. The lower threshold is determined by $\gamma$ where as the upper threshold is determined by $A$. Moreover, the order of magnitude of $\Delta t$ can be estimated as $N^\beta$. The variation of $\beta$ with variation in $A$ and $\gamma$ is shown in Fig.~\ref{figparam} and is found to follow a linear relation which indicates power-law scaling of the order parameter close to the critical transition time during phase transition \cite{achlioptas2009explosive}. Firstly, it is interesting to note that such a linear relation between $A$, $\gamma$ and $\beta$ exists for all $0.01 < St < 1$. Further, the scaling exponents of such a power-law are found to be same for $0.01 < St < 1$, reaffirming the universal behavior of inertial particles with distinct $St$ during clustering.
%\textcolor{red}{@vignesh, praveen check urgently  in what range of St is this claim valid??}

For $St = 0$, the particles are essentially tracer particles and do not form any spatial clusters as they follow the fluid streamlines indefinitely. Therefore, there is no phase transition observed for such particles, and thus $\Delta t \to \infty$ as $St \to 0$. For inertial particles with $St \sim O(1)$, the response time of particles is close to the flow timescale and the particles tend to cluster rapidly. Figure \ref{fig_deltaTwithSt} shows the variation of $\Delta t$ with $St$ which clearly follows a power-law with a scaling exponent $\beta = -0.95\pm 0.01$ with $90\%$ confidence, initially for $St<0.25$. For particles with $0.25\leq St\leq1$, the phase transition time varies in the form $\Delta t \sim e^{(C/St)}$ with the constant $C = 0.18\pm 0.02$ with $90\%$ confidence. This deviation from power-law behaviour could possibly be due to the formation of caustics amongst the particles in this range $0.25\leq St\leq1$ and the variation of  $\Delta t$ is consistent with the rate of caustics formation \cite{samant2021dynamic,nath2021transport}.% We believe that the existence of such a power-law scaling in the range $St<0.25$ and a deviation from it in the range $0.25\leq St\leq1$ is a novel discovery and may be very helpful in building physical intuitions about particle-laden flows such as the rate of clustering of droplets in clouds. 
%\textcolor{red}{@Vignesh: Read the 2020/21 paper which showed the inv. dependence of t and St and write implications of this power law behavior in the above para.}

%For example, the cloud droplet size growth bottle-neck is an unresolved physical problem on how a cloud droplet of a typical size of $10 \mu m$ grows via collision and coalescence to $60 - 100 \mu m$ \cite{grabowski2013growth}. The turbulence in clouds is believed to play a major role in the growth of droplets, by inducing clustering and strong relative velocities between the particles. From literature, the turbulence parameters in typical warm clouds are $u' = 1\ msec^{-1}$, $\epsilon \approx 0.01\ m^2 sec^{-3}$ and $St$ varies between $0.01$ to $2$ for droplet sizes $10 \mu m$ to $60 \mu m$, respectively. Using complex networks on particles, we found that the time taken for the particles to cluster and form a giant component varies as a power law. That is, as particle size increases, the time taken for particles to cluster and interact decreases sharply. More investigation in this regard is advocated and immediate helpfulness of complex networks to study particle-laden flows is thus highlighted.

\section*{DISCUSSION}

We propose the use of complex networks to study the local and global dynamics in particle-laden flows. The local attributes of a network such as the degree of any individual particle captures the local clustering characteristics along its trajectory. On the other hand, the saturation of average network measures reflects the global clustering characteristics of the particles. Further, the distortion of the initial power-law degree distribution reflects the spatial inhomogeneities among particles due to cluster formation. As the connectivity of the proximity-based particle network increases, a giant component emerges in the network. Such an emergence of giant component signifies the formation of particle-clusters in the flow and this process occurs via a continuous phase transition. The network approach reveals a universal behavior of the inertial particles close to the critical time of phase transition evoked by the clustering dynamics in the flow despite the differences in $St$. The phase transition time exhibits a power-law behaviour for $St<0.25$ and an exponential behaviour for $0.25\leq St\leq1$ possibly due to the formation of caustics.

\section*{\label{sec:materials_and_methods}MATERIALS AND METHODS}

\subsection*{Model}\label{Model}

\begin{figure}[h!]
\centering
\includegraphics[width=0.6\textwidth]{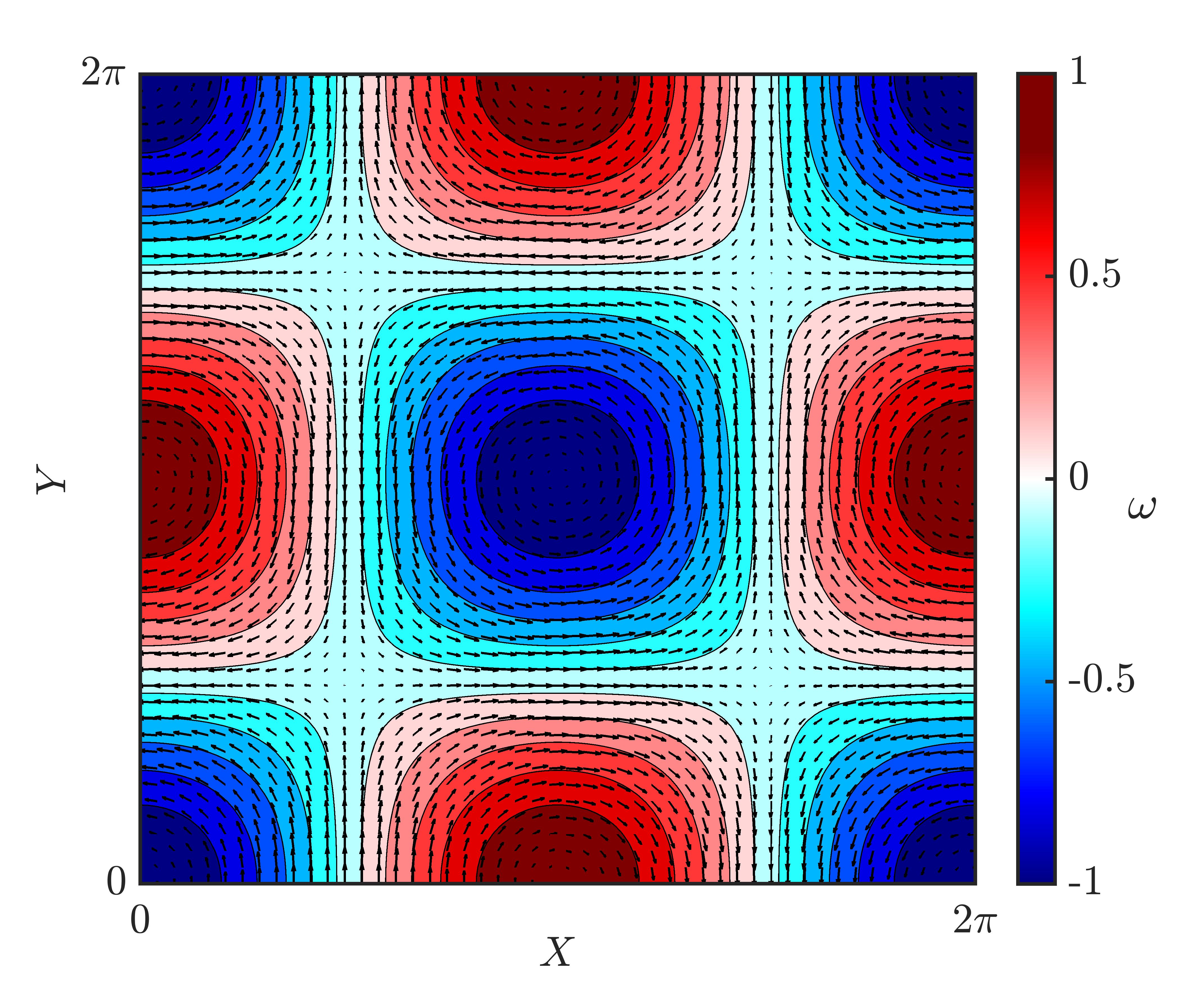}
\caption{ \textbf{Contour plot of the vorticity field of steady 2D Taylor-Green flow overlaid with velocity vectors.} The flow domain consist of 4 vortices with shear regions between them.}
\label{TG_flow}
\end{figure}

In order to understand the clustering process using complex networks, we choose the steady 2-D Taylor-Green flow \cite{taylor1937mechanism,samant2021dynamic} to be our canonical flow (Fig.~\ref{TG_flow}). This simple flow was chosen for our analysis as it has some of the features of turbulent flows such as coexisting regions of vorticity and shear\cite{taylor1937mechanism, maxey1983equation,samant2021dynamic}. 

The velocity profile equations for Taylor-Green flow are as follows: 
\begin{equation}\label{eq1}
u_{f}(x,y) =  V_{0}\left [-cos\left ( {\frac{ 2\pi x}{L_f}} \right )sin\left ( { \frac{2\pi y}{L_f}} \right ) \right ],
\end{equation} 
\begin{equation}\label{eq2}
v_{f}(x,y) =  V_{0}\left [sin\left ( \frac{2\pi x}{L_f} \right ) cos\left ( \frac{2\pi y}{L_f}\right ) \right ]
\end{equation} 
In the above equation, $u_{f}$ and $v_{f}$ are the two components of fluid velocity. Here, $L_f$ and $V_{0}$ are the length of the flow domain and the velocity of the fluid flow. The time scale, $t_{f}$, of the flow is calculated as $t_{f} = L_f/V_{0}$. These aforementioned parameters are used to non-dimensionalize the system of equations in order to minimize the number of input parameters for initiating the simulation. The particles are modeled as point particles having a significantly higher density compared to the surrounding fluid, i.e., $\rho_p/\rho_f =  10^3$. Due to the high density ratio, the contributions of added mass effect, buoyancy and Basset history force were neglected\cite{maxey1983equation}. Therefore, following the Maxey-Riley equation \cite{maxey1983equation}, only the effect of Stokes drag is considered in the particle dynamics. 

%Further, we tested the variation of global network properties of the system, relative to the number of particles considered and found them to be invariant of the number of particles considered if we construct the network as suggested. This has been discussed in detail in the later this section.

We considered $5\times10^3$ particles in our analysis to maintain a sufficiently low volume fraction, to ensure the validity of the unidirectional coupling and to keep the computational time within reasonable limits. In unidirectional coupling, the particle motion is affected by the local fluid structures but the collective motion of the particle does not affect the fluid structures. The dimensionless form of the equations are:
\begin{equation}\label{eq3}
\frac{{d} \vec{x}_{p}(x,y)}{{d} t} = \vec{u}_{p},
\end{equation} 
\begin{equation}\label{eq4}
 \frac{{d} \vec{u}_{p}(x,y)}{{d} t} = \frac{1}{St} \left ( \vec{u_{f}} - \vec{u}_{p} \right ).    
\end{equation}
Here, $\vec{x}_{p}$, $\vec{u_f}$ and $\vec{u_p}$ are the non-dimensional particle position, flow velocity, and particle velocity vectors, respectively. For obtaining different spatial distributions of particles, we perform the analysis for various $St$ values from $0.01$ to $3$. We limit our study in the range $St$ less than $3$, as the assumption of neglecting the added mass effect and Basset history force becomes invalid for $St$ greater than $3$\cite{cencini2006dynamics}.  

At $t$ = 0, the particles are randomly distributed across the flow domain with initial velocities identical to that of the surrounding flow velocity. Then, the particle inertia is activated. We implement a periodic boundary condition across the edges of the flow domain. The simulations are performed up to $100$ non-dimensional units of time with a time step of $10^{-3}$. The simulation results are consistent with previously reported results from the literature \cite{maxey1987motion,crisanti1990passive,samant2021dynamic}. Using the simulated data, we construct complex networks to study the particle clustering process.

%For $St$ = 0.01 particles behave like tracers and not much clustering is observed. % Particle clustering increases as we increase upto $St = 1$, where it reaches maximum. As we further increase $S$t clustering decrasees and particle motion becomes random. 

%For $St$ = 0.1 \& 1, we observe clustering amongst particles with $St$ = 1 showing maximum clustering. For $St$ = 10, particles cluster high inertia and do not respond to the local changes to the flow structures preventing them from clustering.

% \subsection*{Choice of resolution of radial distribution function}\label{RDF_choice}

\subsection*{Effect of number of particles in the flow domain}\label{RDF_choice}

\begin{figure}[h!]
\centering
\includegraphics[width=1\textwidth]{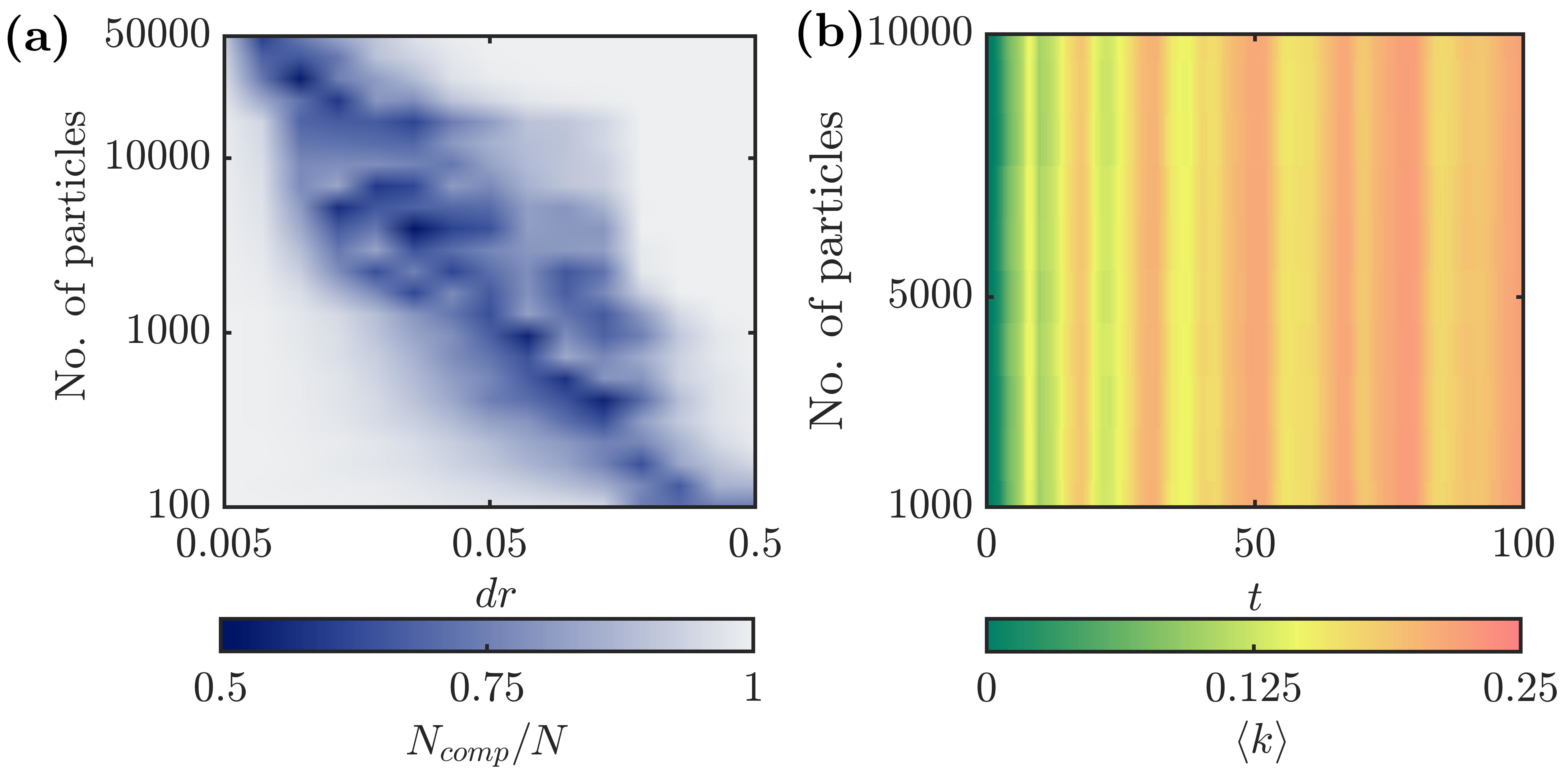}
\caption{\textbf{Optimal spatial resolution ($\boldsymbol{dr}$) for estimating the radial distribution function ($\boldsymbol{g(r)}$) depends on the number of particles in the flow domain ($\boldsymbol{N}$).} (a) Variation of normalized number of components in the network ($N_{comp}/N$) with number of particles in the flow domain ($N$) and the spatial resolution ($dr$) used to evaluate $g(r)$. For a specific value of $N$, to obtain optimal resolution in $g(r)$, we chose $dr$ where $N_{comp}/N$ reaches a minimum. (b) Variation of $\langle k \rangle$ and $N$ for $St = 0.5$ particles. For the optimal $dr$, the dynamics are invariant of $N$.}
\label{N_dr}
\end{figure}
Though $g(r)$ is defined as a continuous function, for estimating it, we need to discretize the space in the vicinity of the particle along the radial direction.
Since, we construct our networks from $g(r)$, an optimal discretization resolution ($dr$) should be selected, depending upon the number of particles in the flow domain ($N$) and the size of the flow domain ($L_f$). Towards this purpose, we analyzed the variation of the normalized number of components ($N_{comp}/N$) of the network constructed from $N$ randomly distributed particles with respect to the variation of $dr$ for a fixed flow domain size $L_f$ (see Fig.~\ref{N_dr}). We fixed the size of the flow domain as $2\pi$ and varied $N$ from 100 to 50000. 

For a specific value of $N$, for a small $dr$ values, $g(r)$ will have sharp peaks at the locations where particles are present and remain zero everywhere else, making it unsuitable for network construction. Hence, the conditions given in Eq.~\ref{eq2} will not be satisfied and the links between particles cannot be established. Thus, the network has a few connections ($N_{comp}/N \approx 1$) (see Fig.~\ref{N_dr}a). For very large $dr$, $N_r/\delta A \approx N/A$, hence $g(r) \approx 1$ becomes incapable of capturing the spatial inhomogeneities, leading to $N_{comp}/N \approx 1$ (see Fig.~\ref{N_dr}a). Between the two extremes, there are intermediate values of $dr$ where $N_{comp}/N$ reaches a minimum. We choose the $dr$ where $N_{comp}/N$ goes below unity and reaches a minimum in our study. At this resolution $dr$, the $g(r)$ calculated is considered optimal for constructing the network and capturing the spatial inhomogeneities. When the network is constructed using this optimal $dr$, the degree distribution of the network at $t$=0 corresponding to $N$ randomly distributed particles follows a power-law behavior. To verify this, we studied the variation of $\langle k \rangle$ with time for $St = 0.5$ particles by varying $N$ from 1000 to 10000 (see Fig.~\ref{N_dr}b). The variation of $\langle k \rangle$ does not depend to the choice of $N$ and therefore indicative of the robustness of this method in analyzing the clustering characteristics in particle-laden flows.

%\textbf{add a gephi plot and degree distribution plot here for reference}. 

\subsection*{Network measures}\label{}

The degree of a node $k_i$ can be expressed in terms of the adjacency matrix as follows,
\begin{equation}\label{eq7}
k_{i} = \sum_{j=1}^N A_{ij}.
\end{equation}
The average degree ($\langle k \rangle$) is calculate as 
\begin{equation}\label{eq9}
\langle k \rangle = \frac{1}{N}\sum_{i=1}^N k_{i}.
\end{equation}
The local clustering coefficient ($C_{i}$)
\begin{equation}\label{eq11}
C_{i} = \frac{2l_i}{k_i(k_i-1)}
%C_{i} = \frac{l_i}{\frac{k_i(k_i-1)}{2}}
\end{equation} 
where $l_i$ is the total number of links between the $k_i$ neighbors of node i.
The average clustering ($\langle C \rangle$) is calculate as 
\begin{equation}\label{eq12}
\langle C \rangle = \frac{1}{N}\sum_{i=1}^N C_{i}.
\end{equation}

%\subsection*{Characteristics of continuous phase transition}\label{}

%Recall the relations used for estimating the phase transition time $\Delta t = t_1 - t_0$, where $t_0$ is time when $C_M \times N \leq N_{max}^\gamma$ and $t_1$ is the time when $C_M \geq \times A$. The $\Delta t$ of the phase transition is estimated based on above equations for $ \gamma \in [0.6, 0.8]$ and $A\in [0.4, 0.8]$. Furthermore, the estimated $\Delta t$ can be approximated as $N^\beta$ \cite{achlioptas2009explosive}. We analyzed the variation of $\beta$ with respect to $\gamma$, $A$ and found the relation to be linear (See Fig.~\ref{LS}a for $St = 0.05$). The linear relation indicates a power-law behavior in the phase transition. Furthermore we found the linear relation to be universal for the all the $St$ analysed in this study. The linear relation for $A = 0.8$ was found to be $\gamma + 2.27\times\beta = 1.19$ (shown in Fig.~\ref{LS}b ).  

%and increases as we move radially outward from the particle. Though $g(r)$ predicts a presence of clustered distribution, it is not in the immediate vicinity of the reference particle. Hence we do not make any connections with our reference particle.

%$\frac{\mathrm{d} g }{\mathrm{d}r} \at[\big]{r=r_c}< 1$.

\section*{Supplementary materials}
movie S1.  Motion of $St = 0.05$ particles in the flow domain with the color representing its corresponding degree in log-scale. \\
movie S2.  Motion of $St = 0.5$ particles in the flow domain with the color representing its corresponding degree in log-scale. \\
movie S3.  Motion of $St = 1$ particles in the flow domain with the color representing its corresponding degree in log-scale. 

\bibliography{scibib}

\bibliographystyle{Science}

\section*{Acknowledgments}
\textbf{Funding:} We gratefully acknowledge the funding from J. C. Bose fellowship (JCB/2018/000034/SSC) from the Science and Engineering Research Board (SERB) of the Department of Science and Technology (DST), Government of India.\\
\textbf{Author contributions:}
%K. S. V. wrote the code. K.S.V and S. T. performed the analysis. All authors contributed towards conceptualizing the research and writing the manuscript.
K. S. V.: conceptualization, developed source code, simulations, analysis, writing and editing original draft; S. T.: conceptualization, validation and analysis of phase transition, writing and editing original draft; P. K.: conceptualization, supervision, writing and editing draft; R. I. S.: conceptualization, supervision, writing and editing draft;

\section*{Declaration of competing interests} 
The authors declare that they have no competing interests.
\section*{Data and materials availability}
All data needed to evaluate the conclusions are present in the paper and the supplementary Materials. Additional data related to this paper will be available upon request to the authors.

%The data used in this paper will be available upon reasonable request to the corresponding co-author.

% Your references go at the end of the main text, and before the
% figures.  For this document we've used BibTeX, the .bib file
% scibib.bib, and the .bst file Science.bst.  The package scicite.sty
% was included to format the reference numbers according to *Science*
% style.

%BibTeX users: After compilation, comment out the following two lines and paste in
% the generated .bbl file. 

%Here you should list the contents of your Supplementary Materials -- below is an example. 
%You should include a list of Supplementary figures, Tables, and any references that appear only in the SM. 
%Note that the reference numbering continues from the main text to the SM.
% In the example below, Refs. 4-10 were cited only in the SM.     

% For your review copy (i.e., the file you initially send in for
% evaluation), you can use the {figure} environment and the
% \includegraphics command to stream your figures into the text, placing
% all figures at the end.  For the final, revised manuscript for
% acceptance and production, however, PostScript or other graphics
% should not be streamed into your compliled file.  Instead, set
% captions as simple paragraphs (with a \noindent tag), setting them
% off from the rest of the text with a \clearpage as shown  below, and
% submit figures as separate files according to the Art Department's
% instructions.

\clearpage

\end{document}